\documentclass[journal]{IEEEtran}

\usepackage[linesnumbered,ruled]{algorithm2e}
\usepackage{algpseudocode}
\usepackage{amsmath}
\usepackage{cite}
\usepackage{graphicx}
\usepackage{footnote}
\usepackage{bbding}
\usepackage{pifont}
\usepackage{wasysym}
\usepackage{amssymb}
\usepackage{multirow}
\usepackage[table,xcdraw]{xcolor}
\usepackage{diagbox}
\usepackage[flushleft]{threeparttable}
\makesavenoteenv{tabular}
\usepackage{tabu}
\usepackage{longtable}
\usepackage{cite}
\usepackage{epstopdf}
\usepackage{color}
\usepackage{enumitem}
\usepackage{booktabs}
\usepackage{subfigure}
\setlist{parsep=0pt,listparindent=\parindent}
\graphicspath{ {image/} }
\usepackage{caption}
\usepackage{amsthm}

\usepackage[utf8]{inputenc}
\ifCLASSINFOpdf
\else
\fi

\usepackage{ragged2e}
\usepackage{verbatim}
\begin{document}
%
\title{MGA: Momentum Gradient Attack on Network}
%
%
%
%


\author{Jinyin~Chen,
        Yixian~Chen,
        Haibin~Zheng,
        Shijing~Shen,
        Shanqing~Yu,
        Dan~Zhang,
        and~Qi~Xuan,~\IEEEmembership{Member,~IEEE}
\thanks{Corresponding author: Shanqing Yu.}
\thanks{J. Chen, Y. Chen, H. Zheng, S. Shen, S. Yu, D. Zhang and Q. Xuan are with the Institute of Cyberspace Security, College of Information Engineering, Zhejiang University of Technology, Hangzhou 310023, China.
(e-mail: \{chenjinyin, 2111803168, 201303080231, 2111803074, yushanqing, danzhang, xuanqi\}@zjut.edu.cn).}}


%
%

\markboth{}%
{Chen \MakeLowercase{\textit{et al.}}: MGA: Momentum Gradient Attack on Network}
\maketitle
%



\IEEEtitleabstractindextext{%
\begin{abstract}
\justifying
The adversarial attack methods based on gradient information can adequately find the perturbations, that is, the combinations of rewired links, thereby reducing the effectiveness of the deep learning model based graph embedding algorithms, but it is also easy to fall into a local optimum. Therefore, this paper proposes a Momentum Gradient Attack (MGA) against the GCN model, which can achieve more aggressive attacks with fewer rewiring links. Compared with directly updating the original network using gradient information, integrating the momentum term into the iterative process can stabilize the updating direction, which makes the model jump out of poor local optimum and enhance the method with stronger transferability. Experiments on node classification and community detection methods based on three well-known network embedding algorithms show that MGA has a better attack effect and transferability.

\justifying
\end{abstract}

}

\maketitle

\IEEEdisplaynontitleabstractindextext

%
\IEEEpeerreviewmaketitle

\ifCLASSOPTIONcompsoc
\IEEEraisesectionheading{\section{Introduction}\label{sec:introduction}}
\else
\section{Introduction}
\label{sec:introduction}
\fi

%
%
%
%

\IEEEPARstart{N}{owadays} many systems can be modeled as graphs, such as social networks~\cite{chen2017interplay,Ding2017TeamGen,xuan2018social}, biological networks~\cite{huang2019integrated,muchowska2019synthesis}, traffic networks~\cite{an2017network,zhang2017bio}. With the help of graph embedding, especially some deep learning models, nodes and links in the network can be represented as low-dimensional vectors, based on which downstream tasks can be easily implemented. Such approaches have achieved great success in various applications such as node classification~\cite{gao2009graph,jian2018toward,ping2018batch}, community detection~\cite{cheng2018novel,govindan2017k,Cheng2018A} and link prediction~\cite{fu2018link,pecli2018automatic,cen2019trust}.

However, it has been found that deep models are vulnerable to adversarial attacks~\cite{goodfellow6572explaining,szegedy2013intriguing,Akhtar2018Threat,tramer2017space,yang2018adversarial}. In the areas such as computer vision~\cite{moosavi2016deepfool,elsayed2018adversarial,cai2018curriculum,duan2019things} and recommendation system~\cite{tang2019adversarial,fang2018poisoning}, the adversarial attacks against deep learning models are considered as a severe security problem, because small perturbations may fool the methods and consequently produce completely wrong output. Recently, the adversarial attacks against network algorithms are also carried out to study the robustness of network algorithms~\cite{bojcheski2018adversarial,dai2018adversarial}, by rewiring a small number of links in the original network. Some even study on using the adversarial attack methods to solve privacy problems, that is, to protect certain sensitive information of the network from being leaked by graph mining methods~\cite{chen2018fast,yu2018target}.

At present, many adversarial attacks on network algorithms directly based on gradient information, i.e., by calculating the partial derivatives of the target loss function to the element of adjacency matrix~\cite{bojcheski2018adversarial,zugner2018adversarial,chen2018fast}. In this way, the generated adversarial network can easily drop into poor local maxima and thus "overfit" the model, leading to lower transferability across models.

Adopting the momentum method~\cite{polyak1964some}, which is a technique to accelerating gradient descent by accumulating a velocity vector in the gradient descent directions, is a widely used method to avoid inadequate local solutions~\cite{bishop1995neural,haykin1994neural,ripley1996pattern}. In computer vision, many studies on the momentum method have been proposed to improve the gradient method~\cite{ruder2016overview,gorodetsky2018gradient,qiao2019efficient,zhang2018energy}. Moreover, it was used in adversarial attacks of image classification algorithms to further improve the success rates of black-box attacks~\cite{dong2018boosting}. However, momentum has not been used in graph adversarial attack. In order to further improve the attack effectiveness of gradient-based adversarial attack against the Graph Convolutional Network (GCN) model, this paper proposes a momentum gradient attack (MGA) that uses the gradient momentum to update the adversarial network. Specifically, we use momentum to produce a new updating direction by adding velocity vector into the gradient of the loss function, which can stabilize the updating direction and make the attack robust to different models. The main contributions of the proposed MGA are as follows.


 \begin{itemize}
\item As a new adversarial attack against the network algorithms, MGA is superior to several state-of-the-art adversarial attacks under the same scale of perturbation, including GradArgmax ~\cite{dai2018adversarial}, RL-S2V ~\cite{dai2018adversarial}, NETTACK ~\cite{zugner2018adversarial}, and Fast Gradient Attack (FGA)~\cite{chen2018fast}.
\item MGA has strong transitivity, and adversarial networks generated based on GCN can also be used to effectively attack other deep models, including DeepWalk~\cite{perozzi2014deepwalk}, node2vec~\cite{grover2016node2vec} and Graph-GAN~\cite{wang2017graphgan}.
\item Compared with FGA~\cite{chen2018fast}, MGA performs better and is more robust in the absence of some structural information of the original network, so it is more practical.
\end{itemize}

The rest of paper is organized as follows. In Sec.~\ref{sec:RW}, we review the existing adversarial attack methods on graph data. In Sec.~\ref{sec:Background}, we briefly introduce the GCN and FGA. In Sec.~\ref{sec:Method}, we introduce MGA method in details, including white-box and black-box adversarial attacks. In Sec.~\ref{sec:Exp}, we empirically evaluate MGA on multiple tasks, and compare the attack performance on several real-world networks. In Sec.~\ref{sec:Conclusion}, we conclude the paper and highlight future research directions.

\section{Related Work}
\label{sec:RW}
Recently, adversarial attack on networks has attracted a lot of attention and many algorithms have been proposed. In particular, for community detection, Nagaraja~\cite{nagaraja2010impact} first introduced a community hiding algorithm, where they added links based on centrality values to change the community structure. Waniek et al.~\cite{waniek2018hiding} proposed a heuristic method named Disconnect Internally, Connect Externally (DICE), which deletes the links within the same communities while adds the links across different communities. Valeria et al.~\cite{fionda2018community} proposed the safeness-based method, which adds and deletes specific links, leading to the maximum increase of safeness. Chen et al.~\cite{chen2019ga} further introduced a community detection attack algorithm based on genetic algorithm, which treated modularity $Q$ as fitness function. For link prediction, Zheleva et al.~\cite{zheleva2008preserving} proposed the link re-identification attack to protect sensitive links from the released data. Link perturbation is a common technique in those early researches that data publisher can randomly modify links on the original network to protect the sensitive links from being identified. Fard et al.~\cite{fard2012limiting} introduced a subgraph wise perturbation in directed networks to randomize the destination of a link within subgraphs so as to protect sensitive links. They further proposed a neighborhood randomization mechanism to probabilistically randomize the destination of a link within a local neighborhood~\cite{fard2015neighborhood}. For link prediction, Yu et al.~\cite{yu2018target} designed evolutionary perturbations based on genetic algorithm and estimation of distribution algorithm to degrade the popular link prediction method Resource Allocation Index (RA). And they found that although the evolutionary perturbations are designed to against RA, it is also effective against other link-prediction methods. For network embedding, Xuan et al.~\cite{xuan2019unsupervised} proposed Euclidean Distance Attack (EDA), a genetic algorithm based network embedding attack method, which focuses on disturbing the Euclidean distance between nodes in the embedding space.

Moreover, due the great success of many graph deep learning frameworks, e.g., Graph Convolutional Networks (GCN), several deep learning based attacks have been proposed recently. Bojchevski and G\"unnemann~\cite{bojcheski2018adversarial} attacked the embedding model by maximizing the random walk model based loss function, and they demonstrated that the generated adversarial network has strong transferability. Dai et al.~\cite{dai2018adversarial} proposed a reinforcement learning based attack, which uses a hierarchical Q-function to reduce its time complexity. Zugner et al.~\cite{zugner2018adversarial} proposed NETTACK, which maximizes the difference in the log-probabilities of the target node before and after the attack. This method not only considers rewiring links, but also modifies nodes' features to generate unnoticeable perturbations. Depending on the gradient information, Chen et al.~\cite{chen2018fast} proposed the Fast Gradient Attack (FGA), which directly uses the gradient of adjacency matrix to iteratively update the adversarial network. By using this method, the embedding of target node can be changed significantly through rewiring a small number of links, so as to greatly affect the downstream algorithms based on embedding. However, although it is relatively fast, FGA easily falls into the poor local optimum since it uses the gradient information directly.

\section{Background}
\label{sec:Background}
\subsection{Double-Layer GCN model}
Since the end-to-end node classification model GCN~\cite{kipf2016semi} has great performance in node classifications, we generate adversarial networks based on gradient information of GCN same as the work in~\cite{zugner2018adversarial,chen2018fast}. Given a network $G$ and the feature of nodes $X$. Denoting $A$ as the adjacency matrix of the network, the forward propagation of GCN model of a single hidden layer then is defined as follow:
\begin{equation}
Y'(A)=f(\bar{A}\sigma(\bar{A}XW_0)W_1),
\label{equ:forward}
\end{equation}
where $\bar{A}=\tilde{D}^{-\frac{1}{2}}\tilde{A}\tilde{D}^{-\frac{1}{2}}$, $\tilde{A}=A+I_N$ is the self-connected form of undirected network $G$ after adding the identity matrix $I_N$ to $A$, $\tilde{D}_{ii}=\sum_j\tilde{A}_{ij}$ is the diagonal degree matrix of $\tilde{A}$; $W_0\in R^{|V|\times H}$ and $W_1\in R^{H\times |F|}$ are the input-to-hidden and hidden-to-output weight matrices, respectively, with the node set $V$, category set $F$ and hidden layer of $H$ feature maps; $f$ and $\sigma$ are the softmax function and Relu active function, respectively.
For semi-supervised multi-label classification on GCN, we define the loss function based on the cross-entropy error over all labeled nodes:
\begin{equation}
L=-\sum_{l=1}^{|V_L|}\sum_{h=1}^{|F|}Y_{lh}\ln(Y'_{lh}(A)),
\label{equ:loss1}
\end{equation}
where $V_L$ is the set of nodes with ground truth labels, $Y$ is the real label matrix with $Y_{lh}=1$ if node $v_l$ belongs to $h$-th category and $Y_{lh}=0$ otherwise, and $Y'(A)$ is the output of the model defined by Eq.~(\ref{equ:forward}).

The weight matrices $W_0$ and $W_1$ are updated based on gradient descent, so that the GCN model is continuously optimized and the node classification performance is steadily improved. As we can see, in Eq.~(\ref{equ:loss1}), the adjacency matrix $A$ is another critical variable in the loss function. Therefore, we can implement network attacks, that is, to reduce the accuracy of nodes classification, by modifying the adjacency matrix A based on its gradient information.
\subsection{Fast Gradient Attack method}
In this subsection, the FGA method~\cite{chen2018fast}, which is a gradient-based adversarial attack method against the GCN model, is briefly introduced. Based on the trained GCN model, the target loss function $L_t$ of target node $v_t$ is defined as Eq.~(\ref{equ:loss2}). It represents the difference between the predicted label and the real one of the target node $v_t$.

\begin{equation}
L_t=-\sum_{k=1}^{|F|}Y_{tk}\ln(Y'_{tk}(A)),
\label{equ:loss2}
\end{equation}

which represents the difference between the predicted label and the real one of the target node $v_t$. The larger value of target loss function corresponds to the worse prediction result. The basic link gradient matrix $\bar{g}$ is represented by Eq.~(\ref{equ:gradient}).

\begin{equation}
\bar{g}_{ij}=\frac{\partial L_t}{\partial A_{ij}}.
\label{equ:gradient}
\end{equation}
where $A_{ij}$ and $\bar{g}_{ij}$ is the element of adjacency matrix and link gradient matrix respectively.

The primary link gradient matrix $\bar{g}$ is symmetrized by Eq.~(\ref{equ:gradient2}) to obtain the initial Link Gradient Network (LGN) $g$ since the adjacent matrix of an undirected network is symmetry.

\begin{equation}
g_{ij}=g_{ji} = \left\{
\begin{array}{ll}
\frac{\bar{g}_{ij}+\bar{g}_{ji}}{2} & i\ne j \\
0 & i = j
\end{array}\right.
\label{equ:gradient2}.
\end{equation}

In initial LGN $g$, each pair of nodes has a positive or negative gradient value. Note that adding/deleting links between pairs of nodes with larger magnitude gradients may cause the target loss function to change rapidly. Therefore, FGA method add links between the pairs of nodes with maximum positive link gradient, and delete links between the pairs of nodes with maximum negative link gradient.

\section{Method}
\label{sec:Method}
\begin{figure*}[!t]
\includegraphics[width=.9\linewidth]{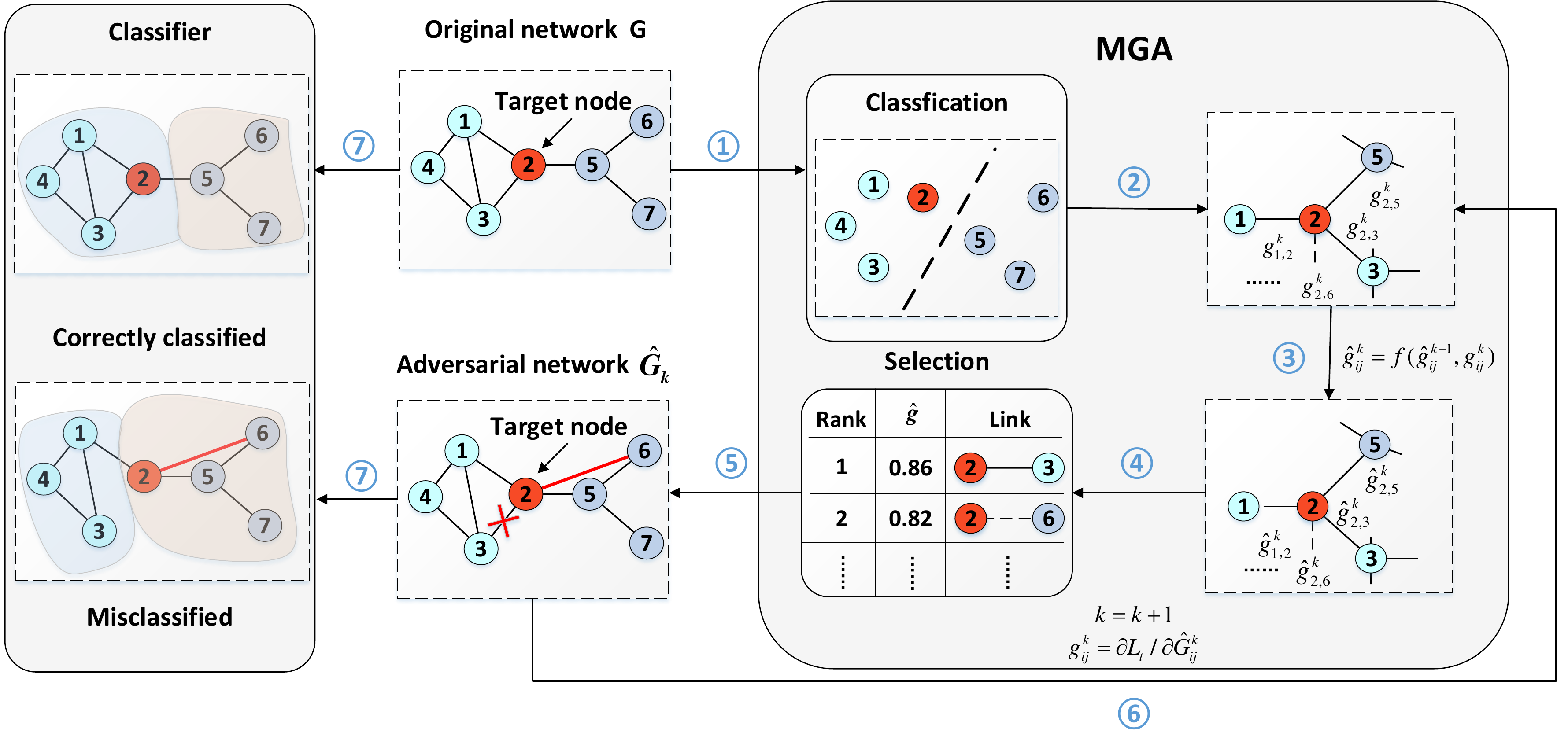}
\centering
\caption{The framework of MGA, which consists of seven steps. First, input original network to train GCN model; second, calculate the gradients of each link for the loss function; third, calculate the momentum based on gradient information; fourth, sort the absolute of momentum; fifth, select the largest one for updating the original network; sixth, recalculate the gradient information based on updated network; finally, gain adversarial network after $K$ times of updates and various node classification algorithms are performed on the adversarial network to validate the attack effect of MGA.}
\label{fig:frame}
\end{figure*}

In this paper, we design MGA as a new attack method to generate adversarial networks based on the trained GCN model. Due to the ability of momentum term in the iterative process to stabilize the updating direction, MGA can successfully fool white-box model (GCN model) as well as black-box models (i.e. DeepWalk, node2vec, GraphGAN, etc). 
In the realization of adversarial attack, we first choose the target nodes, and then generate adversarial networks to hide the selected target nodes under certain network analysis methods. The flowchart of MGA is shown in
Fig~\ref{fig:frame}.
For convenience, the definitions of symbols used in Sec.~\ref{MGA} are briefly summarized in TABLE~\ref{Definition}.

\begin{table}[!t]
\centering
\caption{The definitions of symbols.}
 \label{Definition}
 \resizebox{\linewidth}{!}{
\begin{tabular}{lr}
\hline \hline
Symbol & Definition \\ \hline
   $G=(V,E)$           & the original network with nodes $V$ and links $E$           \\
   $\hat{G}=(V,\hat{E})$       & adversarial network with nodes $V$ and updated links $\hat{E}$           \\
   $A/\hat{A}$           & the adjacency matrix of original network/adversarial network \\
   $L$           & the loss function of the GCN model\\
   $L_t$           & the loss function for target node $v_t$ of the GCN model          \\
   $Y/Y'$          & the label confidence list before/after the attack           \\
   $v_t$         & target node        \\
   $K$           & the maximum number of iterations          \\
   $\hat{G}^k/\hat{A}^k$   & the $k^{th}$ adversarial network/adjacency matrix  \\
   $\bar{g}/g$      & link gradient matrix/network           \\
   $\hat{g}^k$ & the $k^{th}$ momentum of link gradient matrix  \\
  $\mu$ & the decay factor \\
   $\theta$      & the sign function\\

   \hline  \hline
\end{tabular}}
\end{table}

\subsection{Problem Definition}
\label{sec:PD}
We first give the definition of adversarial attack on networks as follows.

\newtheorem{Adversarial attack}{\textbf{\textsc{Definition}}}
 \begin{Adversarial attack}[Adversarial attack]
 Given the original network $G=(V,E)$, where $V$ and $E$ are the set of nodes and links, and the adversarial network $\hat{G}=(V,\hat{E})$  which is generated by rewiring some edges of original network $G$. Generally we can use a network analysis model $f$ to get the analysis results $f(G)$ and $f(\hat{G})$, respectively. The purpose of adversarial attack is to fool the network analysis model $f$ by adversarial network $\hat{G}$, which can be formulated as:
\begin{align}
&\mathop{\arg\max}_{\bar{G}}{\phi(f(G),f(\hat{G}))}, \\
&s.t. \sum_{i<j} |A_{ij}-\hat{A}_{ij}|\le \rho
\label{DEFINITION:2}
\end{align}
where $\rho$ is the size of adversarial perturbation and $\phi$ is a function measuring the difference between $f(G)$ and $f(\hat{G})$.
\end{Adversarial attack}

\subsection{MGA method\label{MGA}}

To generate adversarial network for target node $v_t$ from the original network, after the definition of loss function $L_t$ Eq.~(\ref{equ:loss2}), the network attack methods generate the adversarial network by solving the following constrained optimization problem:
\begin{equation}
\mathop{\arg\max}_{\hat{A}}~L_t(\hat{A}), ~s.t. \sum_{i<j} |A_{ij}-\hat{A}_{ij}|\le \rho,
\label{equ:optimization}
\end{equation}
where $\rho$ is the size of adversarial perturbation.
%

In MGA method, we modify one link during each iteration, and stop the process when the maximum allowed number of modified links is reached. The $k^{th}$ iteration can be described by the following steps.

Firstly with $\hat{A}^0=A$, we can generate the $k^{th}$ LGN $g^{k}$ using the adversarial network adjacency matrix $\hat{A}^{k-1}$ based on Eq.~(\ref{equ:gradient}) and Eq.~(\ref{equ:gradient2}). Then, we can update the $k^{th}$ Link Momentum Network (LMN) $\hat{g}^{k}$ by accumulating the velocity vector in the gradient direction using Eq.~(\ref{equ:momentum}).

\begin{equation}
\hat{g}^{k}_{ij}= \left\{
\begin{array}{ll}
g^{k}_{ij} &  k=1 \\
\mu \cdot \hat{g}^{k-1}_{ij} + \frac{g^{k}_{ij}}{||g^{k}_{ij}||_1} & else
\end{array}\right.
\label{equ:momentum}
\end{equation}

where $\mu$ is the decay factor which in the range of $[0,1]$.

We update the $(k-1)^{th}$ adversarial network by modifying the link $(v_i,v_j)$ of the maximum absolute link momentum in the $k^{th}$ momentum LMN $\hat{g}^{k}$.
\begin{align}
&\hat{A}^k_{ij}=\hat{A}^{k-1}_{ij}+\theta (\hat{g}^k_{ij}),
\label{equ:update} \\
&\mathop{\arg\max}_{(v_i,v_j) \in \Omega}~|\hat{g}^{k}_{ij}|,
\label{equ:update2}
\end{align}
where $\Omega$ is the set of all pairs of nodes except $v_i=v_j$, and $\theta(\hat{g}^k_{ij})$ represents the sign of momentum of the selected pair of nodes $(v_i,v_j)$.

The sign of the momentum is a basic element in LMN which represent the ways of the modifying links, that is, delete or add the links between the selected node pairs. More specially, a positive momentum means to add the links; otherwise, it means to delete the links.

\begin{equation}
\theta (\hat{g}^k_{ij})= \left\{
\begin{array}{ll}
1 & \hat{g}^k_{ij}>0 \\
-1 & else
\end{array}\right.
\label{equ:update3}
\end{equation}

In summary, $\hat{A}^0=A$ and the iterative number $k=1$, initially. Next, LGN $g^k$ is calculated based on Eq.~(\ref{equ:gradient}) and Eq.~(\ref{equ:gradient2}), and LMN $\hat{g}^k$ is generated by Eq.~(\ref{equ:momentum}). Then, the adversarial network is generated iteratively by Eq.~(\ref{equ:update}) and Eq.~(\ref{equ:update2}) and $k=k+1$. When $k=K$, where $K$ is the maximum number of iterations, the algorithm ends and output the final adversarial network. The whole steps of the adversarial network generator is described by the following steps.
\begin{enumerate}
\item Train the GCN model on original network $G$.
\item Initialize number of iterations $k$=1 and the adjacency matrix of the adversarial network by $\hat{A}^0=A$
\item Construct the LGN $g^{k}$ based on the $\hat{A}^{k-1}$ via Eq.~(\ref{equ:gradient}) and Eq.~(\ref{equ:gradient2}).
\item Construct the LMN $\hat{g}^{k}$ based on the $g^{k}$ via Eq.~(\ref{equ:momentum}).
\item Select the pair of nodes $(v_i,v_j)$ of the maximum absolute link momentum in $\hat{g}^k$.
\item Update the adjacency matrix $\hat{A}^k$ via Eq.~(\ref{equ:update})  and Eq.~(\ref{equ:update2})
\item If the maximum number of iterations is not reached ($k<K$), $k=k+1$ and go to step 3); otherwise, go to step 8).
\item Generate the final adversarial network $\hat{G}^K$ based on the adjacency matrix $\hat{A}^K$.
\end{enumerate}

\section{Experiments}
\label{sec:Exp}
\subsection{Experimental Setting}
\label{sec:ES}
Here, we will describe the experimental setting. To testify the effectiveness of MGA, we compare it with some state-of-the-art baseline methods by performing several tasks, including node classification attack, community detection attack and anti-defense test on several well-known datasets. The decay factor $\mu$ used in the experiment is 0.5.

\subsubsection{Datasets}
The Details of three datasets used in the experiment are presented in TABLE~\ref{data sets}.

\begin{table}[!b]
\centering
\caption{The basic statistics of the three network datasets.}
\label{data sets}
\begin{tabular}{c|ccc}
\hline
\hline
Dataset & \#Nodes & \#Links & \#Classes \\ \hline
Pol.Blogs &1,490  & 19,090 & 2  \\
Cora      & 2,708 & 5,429 & 7      \\
Citeseer  & 3,312 & 4,732 & 6      \\\hline \hline
\end{tabular}
\end{table}

\begin{table*}[!t]
\centering
\caption{The attack effects, in terms of ASR and AML, obtained by different attack methods on various network embedding methods and multiple datasets. Here, ASR is obtained by rewiring 20 links.}
\label{table:ua}
\resizebox{\linewidth}{!}{
\begin{tabular}{c|c|ccc|ccc|ccc|ccc}

\hline
\hline
\multirow{3}{*}{Dataset} & \multirow{3}{*}{NEM} & \multicolumn{6}{c|}{ASR (\%)}        & \multicolumn{6}{c}{AML}          \\ \cline{3-14}
                           &                        & \multicolumn{3}{c|}{MGA} & \multicolumn{3}{c|}{FGA} & \multicolumn{3}{c|}{MGA} & \multicolumn{3}{c}{FGA} \\ \cline{3-14}
                           &                        & Unlimited &Direct&Indirect& Unlimited &Direct&Indirect &  Unlimited &Direct&Indirect      & Unlimited &Direct&Indirect    \\ \hline

\multirow{5}{*}{Pol.Blogs} & GCN                    &\textbf{100}& 97.87&34.04  &87.87& 85.74& 25.53     &\textbf{3.23}& 3.74&16.09  &8.42& 8.82& 17.61\\

                           & DeepWalk               &\textbf{97.96}& 91.67&6.25  &84.26& 81.66& 6.25      &\textbf{6.76}& 7.19&19.11  &9.84& 10.93& 19.20   \\

                           & node2vec               &\textbf{97.96}& 91.67&5.15  &84.34& 81.83& 0.00      &\textbf{6.71}& 7.08&19.30  &9.72& 10.16& 20.00  \\

                           & GraphGAN               &\textbf{98.63}& 89.58&5.48  &81.21& 80.24& 0.00      &\textbf{6.72}& 7.15&19.16  & 9.41& 11.02& 20.00   \\ \cline{2-14}

                           & Average                &\textbf{98.64}& 92.70&12.73  & 84.42&  82.37& 9.95      &\textbf{5.86}& 6.29&18.42  & 9.35&10.23& 19.20      \\
                           \hline
\multirow{5}{*}{Cora}      & GCN                    &\textbf{100}& \textbf{100}&84.38  &\textbf{100}& \textbf{100}& 88.28   &\textbf{1.74}& 1.85&8.91  &2.54& 3.21& 6.77  \\

                           & DeepWalk               &\textbf{100}& \textbf{100}&85.98  &\textbf{100}& 97.22& 81.55        &3.27& \textbf{3.03}&9.01  &5.61& 6.27& 10.59   \\

                           & node2vec               &\textbf{100}& \textbf{100}&85.51  &\textbf{100}& \textbf{100}& 84.00   &3.14& \textbf{3.10}&8.71  &5.66 &5.58& 9.52 \\

                           & GraphGAN               &\textbf{100}& \textbf{100}&83.80  &\textbf{100}& 96.00 &84.62        &\textbf{3.17}& 3.22&8.70  &5.65& 6.40& 11.02  \\ \cline{2-14}

                           & Average                &\textbf{100}& \textbf{100}&77.75  & 100&98.31& 84.61               &2.83& \textbf{2.80}&7.89  & 4.87&5.37& 9.48     \\
                           \hline
\multirow{5}{*}{Citeseer}  & GCN                    &\textbf{100}& \textbf{100}&94.23 & \textbf{100}& \textbf{100}& 91.36         &2.53& \textbf{2.40}&7.51  &3.52& 3.88& 7.69 \\

                           & DeepWalk               &\textbf{100}& \textbf{100}&84.48  &\textbf{100}  &\textbf{100}& \textbf{100}  &5.25& \textbf{5.17}&7.64  &5.68 &6.06 &7.76 \\

                           & node2vec               &\textbf{100}& \textbf{100}&85.34&\textbf{100}& \textbf{100}& 98.21          &5.09& \textbf{4.97}&7.69  &5.62& 6.50& 7.75 \\

                           & GraphGAN               &\textbf{100}& \textbf{100}&84.02  &  \textbf{100}& 97.89& 93.15          &\textbf{5.04}& 5.55&7.98  &5.91& 6.67& 8.18 \\ \cline{2-14}

                           & Average                &\textbf{100}& \textbf{100}&87.02  & 100&99.47& 95.68                  &\textbf{4.48}& 4.52&7.71  & 5.18&5.78& 7.85      \\ \hline \hline
\end{tabular}
}
\end{table*}

\begin{itemize}
\item \textbf{Pol.Blogs:} This dataset is compiled by Adamic and Glance~\cite{adamic2005political}, which show the political leaning of blog directories. The blogs are divided into two classes and the links between blogs were automatically extracted from the front pages of the blogs. It contains 1,490 blogs and 19,090 links in total.
\item \textbf{Cora:} This dataset is a typical paper citation network in the computer science domain with the papers divided into seven classes. It contains 2,708 paper and 5,429 links in total~\cite{Mccallum2000Automating}.
\item \textbf{Citeseer:} This dataset consists of 3,312 scientific publications classified into six classes. The citation network consists of 4,732 citation links which represents the citation relationships.
\end{itemize}

\subsubsection{Baselines}
Note that our experimental setting is the same as FGA method~\cite{chen2018fast}. So in the Sec.~\ref{sec:NCA}, we only show the comparison results with FGA algorithm, and FGA algorithm has shown that its effect is better than other algorithms, including NETTACK~\cite{zugner2018adversarial}, RL-S2V~\cite{dai2018adversarial} and GradArgmax~\cite{dai2018adversarial}.
\begin{itemize}
\item \textbf{GradArgmax}~\cite{dai2018adversarial}.  GradArgmax greedily select the links that are most likely to cause the change to the loss function based on gradient information.
\item \textbf{RL-S2V}~\cite{dai2018adversarial}. RL-S2V is a reinforcement learning based attack method which decomposes the quadratic action space by hierarchical method. It use Q-learning to learn the attack procedure which parameterized by S2V~\cite{dai2016discriminative}. We set the iteration number $\Delta=200$ for this method.
\item \textbf{NETTACK}~\cite{zugner2018adversarial}. NETTACK  iteratively select the links which can maximize the changes of the surrogate model's results before and after the attack. Note that it only rewires links which can preserve the graph' s degree distribution and feature co-occurrences.
\item \textbf{FGA}~\cite{chen2018fast}. FGA directly uses the gradient of adjacency matrix to generates adversarial network iteratively. In each iteration, it selects candidate node piars with maximum gradient to update the adversarial network.
\end{itemize}

\subsubsection{Performance metrics}
In order to compare the transferability of different adversarial attack methods, we use the generated adversarial networks to attack other network embedding approaches including DeepWalk~\cite{perozzi2014deepwalk}, Node2vec~\cite{grover2016node2vec} and GraphGAN~\cite{wang2017graphgan}. And we also use the following two metrics to compare the effectiveness of these attack methods.

\begin{itemize}
\item \textbf{ASR}. The average success rate is the ratio of misclassification of target nodes by changing no more than $\rho$ links. In the experiment, the perturbation size $\rho$ is varied from 1 to 20. The higher ASR corresponds to the better attack effect.
\item \textbf{AML}. The average number of rewiring links to successfully misclassify a target node. In the experiment, the maximum number of rewiring links is 20, i.e. , when a target node can't be misclassified by rewiring 20 links, we set the number as 20. The lower AML corresponds to the better attack effect.
\end{itemize}

\subsection{Node Classification Attack}
\label{sec:NCA}
Three node classification attack methods are carried out in this subsection. For each method, we do uniform attack, hub-nodes attack and bridge-nodes attack, respectively.

\begin{itemize}
\item \textbf{Direct attack:} The attack is directed at the target node; that is, the redistributed link is limited to the neighbor link of the target node in the network.
\item \textbf{Indirect attack:} The attack is located on a link that is not connected to the target node to make the attack more covert.
\item \textbf{Unlimited attack:} Generate perturbations without considering whether they are around the target nodes or not.
\end{itemize}

\begin{figure*}[!t]
\includegraphics[width=\linewidth]{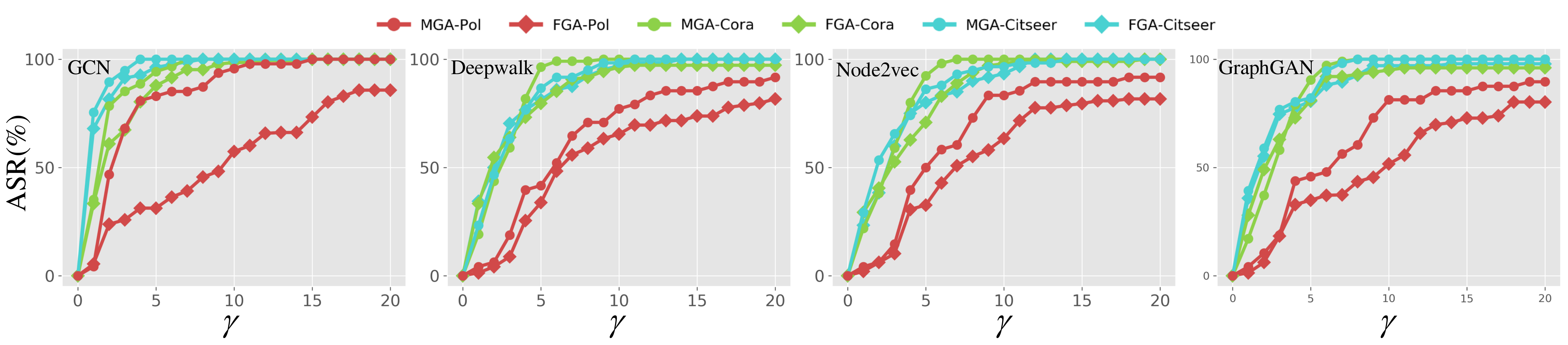}
\centering
\caption{ASR of different attack methods of perturbation size $\gamma$ for different dataset.}
\label{fig:result}
\end{figure*}

We randomly choose 10\% labeled nodes as training and validation sets, respectively. The rest 80\% nodes are used for testing. For all the network embedding methods, the vector dimension is set to 128, the number of walks per node is set to 10, the length of walk is set to 80 and the size of context window is set to 10. The nodes that may be misclassified by the GCN model or logistic regression classifier without attack will not be considered as the target nodes.

\subsubsection{uniform attack}
\label{sec:UA}
The uniform attack experiments, which select 20 nodes from each category as the target nodes, were carried out to verify the general attacking effect of the strategies. The results are shown in Tables~\ref{table:ua} and Figs.~\ref{fig:result}. It can be seen that, in most cases, whether on ASR or AML, MGA achieves better attack results than FGA. Interestingly, such superiority is even more distinct on Pol.Blogs dataset, which has more links and larger attack solutions space.

Attacking GCN will get better attack performance than attacking other network embedding algorithms, since MGA is developed based on GCN. However, when attacking other network embedding algorithms, the performance of MGA is also better than other attack methods,  which proves that MGA has stronger transferability. Even in indirect attacks, MGA performed quite well on the Cora and Citeseer datasets, with ASR exceeding 80\% in most cases.


\subsubsection{Hub-node attack}
\label{sec:HA}
There are many indicators of concentration that reflect the importance of nodes in the network, including degree centrality, closeness, betweenness~\cite{freeman1978centrality}. In this subsection, we select 40 nodes with highest degree centrality~\cite{yustiawan2015degree} as hub nodes and consider them as target nodes. The results presented in Tables~\ref{table:ha} show that, MGA still outperforms other methods, and unlimited MGA outperforms direct or indirect MGA in most cases. Besides, due to the high degree centrality of target nodes, compared to the uniform attack shown in Tables~\ref{table:ua}, the effectiveness of all the attack methods are reduced, especially on the Pol.Blogs dataset whose hub-nodes have higher degree value than other datasets. In addition, we can also find that the transferability of indirected MGA decreases most when attacking hub-node, which may because the target nodes have more neighborhood make their embeddings much more robust and in different models, links far away from hub nodes influence those nodes is different.

\subsubsection{Bridge-node attack}
\label{sec:BA}
In each network, 40 nodes with the highest betweenness centrality~\cite{shimbel1953structural,bavelas1948mathematical}, were selected as bridge nodes. Different from the degree centrality, the betweenness measures the degree to which a node is located on the shortest path between two other nodes~\cite{opsahl2010node,borgatti2005centrality}, and consequently considers the global network structure. The attack results are presented in Table~\ref{table:ba}. Similar to uniform attacks and hub-node attacks, MGA performs better than other methods in the unlimited and directed attacks, and all methods perform best in the unlimited attacks. Meanwhile, it is found that it is easier to attack the bridge nodes than the hub nodes, which may be because the degree of the bridge nodes is much smaller than the hub nodes.

\begin{figure}[!t]
\includegraphics[width=.9\linewidth]{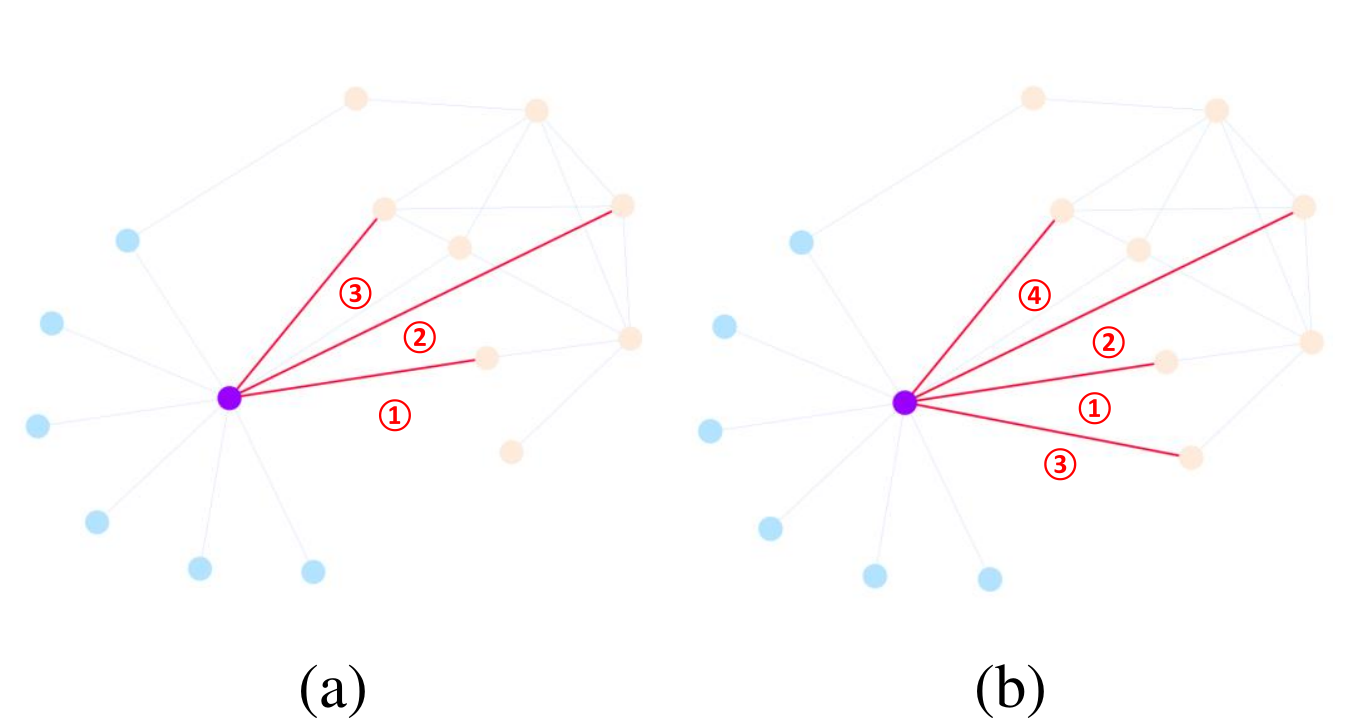}
\centering
\caption{The visualization of target node attack obtained by MGA and FGA. Nodes of the same color belong to the same class and the purple node represent the target node. Note that the red links are the added links and the numbers indicate the order of it. (a) MGA, (b)FGA.}
\label{fig:example}
\end{figure}

Overall, by comparing the results in Table~\ref{table:ua}-Table~\ref{table:ba}, we can find that MGA is useful in attacking different types of nodes, which indicates  that integrating momentum terms into the iterative process can enhance the attack effect and transferability. Taking a target node as an example, the visualization results of the MGA and FGA are shown in Fig~\ref{fig:example}. From Fig~\ref{fig:example}, we can see that the added links of MGA and FGA are the same in the first two iterations, but results of the third iteration are quite different. MGA can attack successfully after three iterations, while FGA needs four iterations to succeed. It shows that MGA finds the optimal solution faster by using momentum in gradient attack.

\begin{table*}[!t]
\centering
\caption{The hub-node attack effects, in terms of ASR and AML, obtained by different attack methods on various network embedding methods and multiple datasets. Here, ASR is obtained by rewiring 20 links.}
\label{table:ha}
\resizebox{\linewidth}{!}{
\begin{tabular}{c|c|ccc|ccc|ccc|ccc}

\hline
\hline
\multirow{3}{*}{Dataset} & \multirow{3}{*}{NEM} & \multicolumn{6}{c|}{ASR (\%)}        & \multicolumn{6}{c}{AML}          \\ \cline{3-14}
                           &                        & \multicolumn{3}{c|}{MGA} & \multicolumn{3}{c|}{FGA} & \multicolumn{3}{c|}{MGA} & \multicolumn{3}{c}{FGA} \\ \cline{3-14}
                           &                        & Unlimited &Direct&Indirect& Unlimited &Direct&Indirect &  Unlimited &Direct&Indirect      & Unlimited &Direct&Indirect    \\ \hline

\multirow{5}{*}{Pol.Blogs} & GCN                    &\textbf{57.5}& 55.0&5.00  &50.00& 45.00& 5.00       &\textbf{14.25}& 14.45&19.20  &14.78& 15.28& 19.70\\
                           & DeepWalk               &47.50& 50.00&2.50  &\textbf{53.85}& 50.00& 2.50      &16.12& \textbf{15.95}&19.55  &16.00& 16.82& 19.77 \\

                           & node2vec               &\textbf{48.75}& 47.50&5.00  &32.50& 45.00& 5.00      &\textbf{17.14}& 17.27&19.23  &17.62& 17.95& 19.20\\

                           & GraphGAN               &\textbf{49.17}& 40.00&2.50  &25.00& 17.50& 2.50      &\textbf{17.07}& 17.27&19.36  &18.68& 18.77& 19.52 \\ \cline{2-14}

                           & Average                &\textbf{50.73}& 48.13&3.75  &40.34&39.38&3.75        &\textbf{16.15}& 16.24&19.34  & 16.77&17.21& 19.55  \\
                           \hline
\multirow{5}{*}{Cora}      & GCN               &\textbf{96.88}& 92.86&55.70  &88.90& 87.18& 54.54         &\textbf{4.88}&4.90&11.15  &6.48& 6.62& 14.37\\

                           & DeepWalk               &\textbf{93.90}& 87.50 &47.22  &85.63& 80.50& 43.75   &\textbf{6.61}& 7.75&13.29  &7.57& 7.73& 13.70 \\

                           & node2vec               &\textbf{91.67}& 91.18&48.39  &81.82& 80.34& 41.96    &\textbf{6.92}& 6.74&12.95  &7.22& 7.52& 13.05\\

                           & GraphGAN               &\textbf{93.75}& 87.50&53.13  &84.36& 82.57& 48.74    &\textbf{7.44}& 7.97&12.19  &8.00& 8.15& 12.98\\ \cline{2-14}

                           & Average                &\textbf{94.05}& 89.76&58.28  & 85.18&82.65& 47.25   &\textbf{6.46}& 6.84&11.84  & 7.32&7.51& 13.53  \\
                           \hline
\multirow{5}{*}{Citeseer}  & GCN                    &\textbf{100}& \textbf{100}&74.29 &\textbf{100}& \textbf{100}& 65.71  &\textbf{1.94}& 2.49&6.54  &5.11& 5.23& 9.77   \\

                           & DeepWalk               &\textbf{94.44}& 91.89&18.92  &88.89& 86.89& 8.33    &\textbf{9.81}& 11.62&18.51  &10.67& 12.14& 18.89 \\

                           & node2vec               &\textbf{93.15}& 90.41&17.81  &86.11& 88.89& 13.51   &\textbf{9.74}& 11.59&18.56  &13.08& 12.14& 18.03\\

                           & GraphGAN               &\textbf{91.74}& 89.91&18.02  &89.19& 88.89& 5.56     &\textbf{10.16}& 11.50&18.20  &13.47& 11.95& 19.17\\ \cline{2-14}
                           & Average                &\textbf{94.83}& 93.05&32.26  & 91.05&91.17& 23.28    &\textbf{7.91}& 9.30&15.45  &10.58&10.37& 16.47    \\ \hline \hline
\end{tabular}
}
\end{table*}

\begin{table*}[!t]
\centering
\caption{The bridge-node attack effects, in terms of ASR and AML, obtained by different attack methods on various network embedding methods and multiple datasets. Here, ASR is obtained by rewiring 20 links.}
\label{table:ba}
\resizebox{\linewidth}{!}{
\begin{tabular}{c|c|ccc|ccc|ccc|ccc}

\hline
\hline
\multirow{3}{*}{Dataset} & \multirow{3}{*}{NEM} & \multicolumn{6}{c|}{ASR (\%)}        & \multicolumn{6}{c}{AML}          \\ \cline{3-14}
                           &                        & \multicolumn{3}{c|}{MGA} & \multicolumn{3}{c|}{FGA} & \multicolumn{3}{c|}{MGA} & \multicolumn{3}{c}{FGA} \\ \cline{3-14}
                           &                        & Unlimited &Direct&Indirect& Unlimited &Direct&Indirect &  Unlimited &Direct&Indirect      & Unlimited &Direct&Indirect    \\ \hline

\multirow{5}{*}{Pol.Blogs} & GCN                    &\textbf{63.16}& 57.9&18.42  &62.16&54.05&5.26   &\textbf{13.71}& 13.92&17.97   &14.65  &14.95  &19.29    \\
                           & DeepWalk               &41.54& \textbf{44.74}&7.84  &36.84&36.84&7.89   &16.97 & \textbf{16.84}&18.83 &17.08  &17.61  &18.82  \\
                           & node2vec               &\textbf{56.76}& 52.11&5.41  &50.00 &52.63&5.26   &\textbf{15.89}& 16.95&19.06  & 16.21 & 16.03 & 19.13\\
                           & GraphGAN               &\textbf{39.47}& 36.84&5.26  &30.77 &25.79&5.26   &\textbf{14.13}& 17.16&19.37  &17.92  & 18.25 & 19.37 \\ \cline{2-14}
                           & Average                &\textbf{50.23}& 47.90&9.23  &44.94 &42.33&5.92   &\textbf{15.18}& 16.22&18.81   &16.47 &16.71  &19.15  \\
                           \hline
\multirow{5}{*}{Cora}      & GCN                &\textbf{96.97}& 93.33&78.57  &92.59&88.89&55.56          &\textbf{3.25}& 4.11&9.14 & 4.63&4.78&10.93 \\
                           & DeepWalk           &\textbf{93.75}& 90.32&41.16  &\textbf{93.75}&90.32&28.12  &\textbf{7.31}& 8.55&14.52 & 7.91&8.03&16.03   \\
                           & node2vec           &\textbf{93.75}& 90.91&45.31  &\textbf{93.75}&86.67&40.00   &\textbf{6.84}& 8.18&14.14 & 8.25&9.03&14.46 \\
                           & GraphGAN           &\textbf{93.25}& 90.32&42.71  &90.62&90.00&33.33            &\textbf{7.93}& 8.48&14.14 & 8.69&8.87&15.97   \\ \cline{2-14}
                           & Average             &\textbf{94.43}& 91.22&51.94  &92.68&88.97&39.25        &\textbf{6.33}& 7.33&12.99& 7.37&7.68& 14.35   \\
                           \hline
\multirow{5}{*}{Citeseer}  & GCN                    &\textbf{100}& \textbf{100}&93.75  &96.97&96.97&66.67   &\textbf{2.03}& \textbf{2.03}&5.50  &4.94&5.03 &11.61   \\
                           & DeepWalk               &\textbf{96.77}& 93.33&45.16  &93.33&93.33&20.69       &\textbf{6.90}& 8.10&15.65  &8.80& 9.33& 18.76  \\
                           & node2vec               &\textbf{95.00}& 93.55&42.62  &93.33&90.32&26.67     &\textbf{6.97}& 7.87&16.39  &9.63& 9.06& 16.27 \\
                           & GraphGAN               &\textbf{93.41}& 90.62&41.76  &93.55&93.33&19.35      &\textbf{6.80}& 7.38&16.62 &  8.97& 8.33& 17.87 \\
                           \cline{2-14}
                           & Average                &\textbf{96.30}& 94.38&55.82  &94.30&93.49&33.35  &\textbf{5.68}& 6.35&13.54  &8.09&7.94&16.13  \\ \hline \hline
\end{tabular}
}
\end{table*}

\subsection{Selected Links Analysis}
\label{sec:SLA}

\begin{center}
\begin{figure*}[!t]
\includegraphics[width=\linewidth]{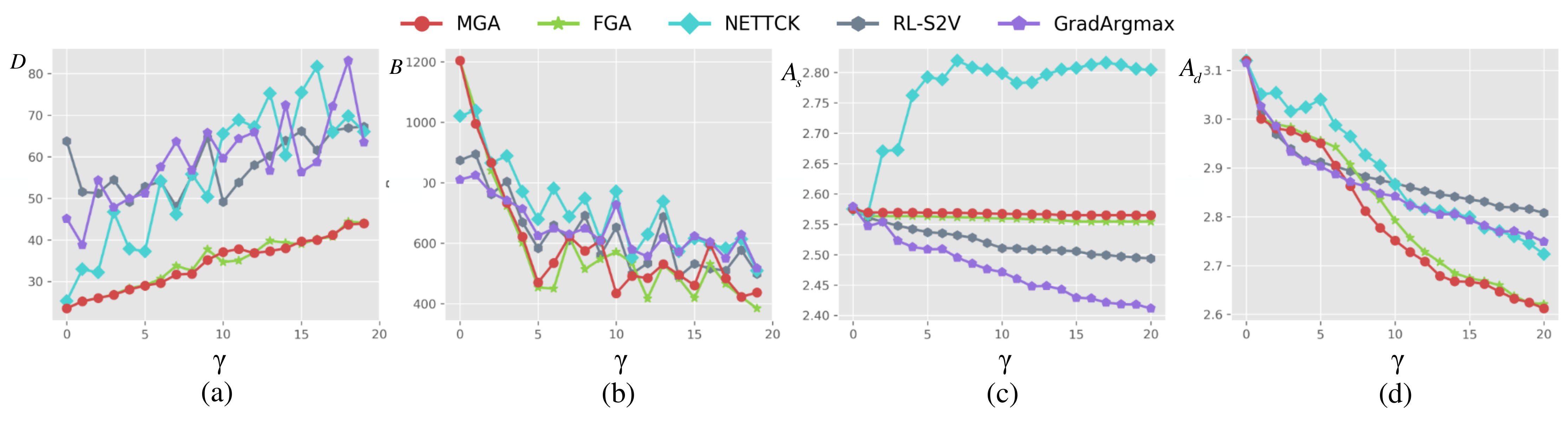}
\centering
\caption{The information of selected links. (a) Node degree; (b) Betweenness; (c) The average distance of target nodes to the same kind of nodes; (d) The average distance of target nodes to the different kinds of nodes.}
\label{fig:Analysis}
\end{figure*}
\end{center}

In this section, we will explore the reason why the links selected by MGA have a greater impact on node classification compared to others. We calculate the four metrics value to evaluate the selected links. For the $\gamma$-th selected links, we calculate the sum of the its related node degree $D(\gamma)$ and its's betweenness $B(\gamma)$. Further more, we also calculate the average distance $A_s(\gamma)$ (resp., $A_d(\gamma)$), which denote the average distance between target node and the labeled nodes which belong to the same class(resp., different classes) after update the $\gamma$ links. More specifically, these metrics are defined as follows.

\begin{equation}
	D(\gamma) =\sum^{E_{\gamma}}_{e=(u,v)}\frac{d_u+d_v}{n_t},
\end{equation}
where $E_{\gamma}=\{e_{\gamma}^{1},e_\gamma^{2},\cdots,e^{n_t}_{\gamma}\}$ is the set of $\gamma$-th selected links for $n_t$ target nodes. Further more $d_u$ and $d_v$ is the degree of node $u$ and $v$ before update those links respectively.

\begin{eqnarray}
    B(\gamma) = \sum_{e \in E_{\gamma}}\sum_{u,v \in V}\frac{\sigma(u,v|e)/\sigma(u,v)}{n_t},
\label{mu-3}
\end{eqnarray}
where $V$ is the node set in the graph, $\sigma(u,v)$ represents the total number of shortest paths between nodes $u$ and $v$, and $\sigma(u,v|e)$ represents the number of shortest paths through the link $e$.

\begin{eqnarray}
A_s(\gamma) =\sum_{v_t\in T}\frac{\sum_{u\in C_t}D_{\gamma}(u,v_t)/n_{C_t}}{n_t},
\label{me-same}
\end{eqnarray}
\begin{eqnarray}
A_d(\gamma) =\sum_{v_t\in T}\frac{\sum_{u\notin C_t}D_{\gamma}(u,v_t)/\bar{n}_{C_t}}{n_t},
\label{me-diff}
\end{eqnarray}

where $D(u,v_t)$ is the distance between target node $v_t$ and node $u$ after update the $\gamma$ links, and $C_t$ is the class which target node $v_t$ belongs to. Furthermore, the number of nodes which belong or not belong  to class $C_t$ is $n_{C_t}$ and $\bar{n}_{C_t}$ respectively.

From Fig.~\ref{fig:Analysis}, We can find the average betweenness $B$ can't be used to judge the result of node classification, but we can generally conclude that the lower the average degree $D$, the better the result of classification will be. This may be because modifying the links on the nodes with small degree value has more influence on those nodes than modifying links on nodes with large degree value. The NETTACK algorithm significantly increases $A_s$ and decreases $A_d$, while MGA, FGA, RL-S2V and GradArgmax don't increase $A_s$.  This is because those methods rarely delete links. We can also find that MGA and FGA both decreases the maximum amplitude $A_d$ as much as possible and ensure that $A_s$ does not decrease much. Considering that the effects of MGA and FGA are better than NETTACK, we believe that the priority of lowering $A_d$ is greater than increasing $A_s$.

\subsection{Community Deception}
\label{sec:CD}
Community detection is aim to find community structure by maximizing cluster quality measures such as modularity~\cite{newman2004finding,newman2016community}. Similar to node classification, community detection also labels each node, but most of them are unsupervised. In this subsection, experiments are performed to study the effect of MGA on community detection methods. Community deception~\cite{nagaraja2010impact,waniek2018hiding,fionda2018community} is strive to hide target community from being discovered by community detection algorithms. We first use attack methods to generate the adversarial networks and then use two community detection algorithms and a combination of network embedding methods and K-means to detect the communities. The two community detection algorithms used in this subsection are introduced as follows
\begin{itemize}
\item \textbf{LPA~\cite{raghavan2007near}}. Firstly each vertex has its own label and then update its label by choosing the most frequent label of its neighbor until the labels don't change. This algorithm has randomness which runs in time $O(E)$.
\item \textbf{Louvain~\cite{blondel2008fast}}. A modularity based algorithm which can generate a hierarchical community structure by compressing the communities as new nodes and its runs in time $O(\left|V\right|\log\left|V\right|)$.
\end{itemize}

The datasets we used is described as follow.
\begin{itemize}
\item \textbf{The USA political books~\cite{newman2006modularity}:} The network represents co-purchasing of US politics books around the time of the 2004 presidential election. The network has 105 books and 441 links in total, and the books are classified as conservative, liberal, and neutral.
\item \textbf{Dolphin social network~\cite{lusseau2003bottlenose}:} The social network represents the common interactions of 62 dolphins which live off Doubtful Sound, New Zealand. The network has 159 links and can be divided into two parts.
\end{itemize}

\begin{center}
\begin{figure}[!t]
\includegraphics[width=\linewidth]{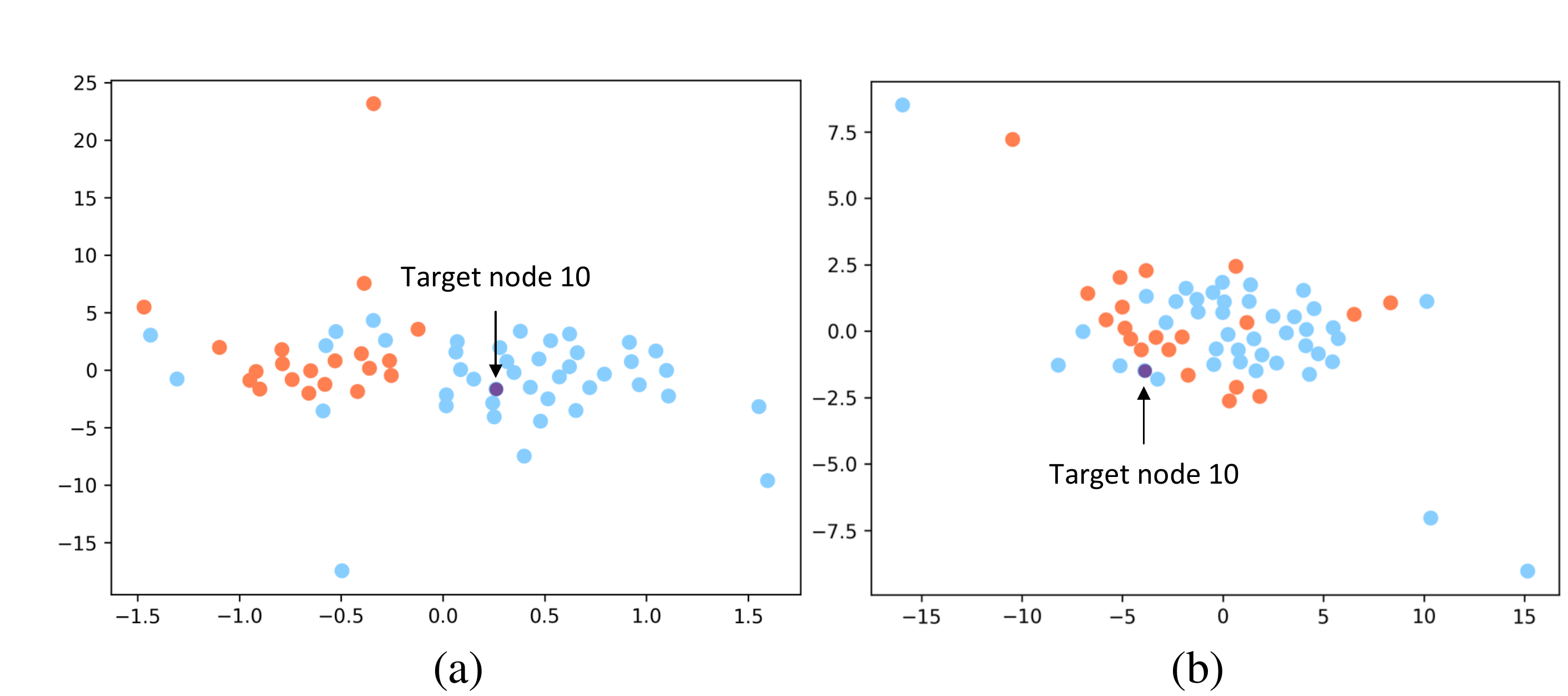}
\centering
\caption{The visualization of target node attack obtained by MGA, utilizing t-SNE algorithm, on Dolphin network. The purple node represent the target node 10, which originally belongs to the blue community. After the attack, it belongs to the red community. Except the target node, the nodes of same color still belongs to the same community detected by the fast greedy algorithm~\cite{newman2004fast}. (a) before attack; (b) after attack.}
\label{fig:TSNE}
\end{figure}
\end{center}

In this subsection, we randomly choose 10\% nodes from each community as training and validation sets respectively. The rest 80\% nodes are used for testing. For all the network embedding methods, the dimension of embedding vector is set to 20 and other settings are consistent with Sec.~\ref{sec:NCA}. As an example, we visualize the attack effect of MGA on the Dolphin network using node2vec and t-SNE ~\cite{maaten2008visualizing}, as shown in Fig.~\ref{fig:TSNE}. From Fig.~\ref{fig:TSNE} we can see that after the attack the target nodes which originally belong to the blue community are more closer to the red community than blue community, while the other nodes still show a more distinct community distribution.

Table~\ref{table:cd} shows the results of each algorithm on community deception, from which we can see that MGA perform better than other attack methods, suggesting that MGA is effective for attacking community detection algorithms. Besides, the effect of each algorithm on Pol.Book is slightly lower than that on dolphins data set, suggesting that community deception is still affected by the degree of target nodes.

\begin{table*}[!t]
\centering
\caption{The attack effects on community detection, in terms of ASR and AML, obtained by different attack methods on various network embedding methods and multiple datasets. Here, ASR is obtained by changing 20 links.}
\label{table:cd}
\resizebox{\linewidth}{!}{
\begin{tabular}{c|c|ccc|ccc|ccc|ccc}

\hline
\hline
 \multirow{3}{*}{Dataset} & \multirow{3}{*}{NEM} & \multicolumn{6}{c|}{ASR (\%)}        & \multicolumn{6}{c}{AML}          \\ \cline{3-14}
                           &                        & \multicolumn{3}{c|}{MGA} & \multicolumn{3}{c|}{FGA} & \multicolumn{3}{c|}{MGA} & \multicolumn{3}{c}{FGA} \\ \cline{3-14}
                           &                        & Unlimited &Direct&Indirect& Unlimited &Direct&Indirect &  Unlimited &Direct&Indirect      & Unlimited &Direct&Indirect    \\ \hline
 \multirow{6}{*}{PloBook}
                           & DeepWalk   & \textbf{100}&98.55&56.06  &93.55& 88.64& 22.95&    \textbf{5.33}&7.28&13.83  &7.84& 8.70& 18.33    \\

                           & node2vec   & \textbf{100}&97.06&69.23   &92.06& 90.26& 19.05&   \textbf{5.12}&7.41&12.55  &8.68& 9.01& 18.41  \\

                           & GraphGAN   & \textbf{100}&95.65&66.67   &93.65& 89.21& 17.46&    \textbf{5.31}&7.54&13.20  &8.32& 8.34& 18.63  \\

                           & LPA        & \textbf{100}&97.37&60.53  &96.37 &95.36 &24.35  &   \textbf{5.24}&5.31&13.42 &5.94 &6.03 &18.21       \\
                           & Louvain    & \textbf{100}&94.74&55.26   &98.38&94.12&21.10  &     \textbf{5.86}&6.05&13.62  &5.91&6.39 &18.38      \\ \cline{2-14}
                           & Average    & \textbf{100}&96.67&61.55   &94.80&91.52 &20.98&       \textbf{5.37}&6.72&13.32  &7.34&7.69 &18.39     \\ \hline
 \multirow{6}{*}{Dolphin}
                           & DeepWalk   & \textbf{100}&\textbf{100}&72.22   &100 &100 &68.18   &   \textbf{3.52}&4.22&13.21  &3.64& 4.39& 13.45\\

                           & node2vec   & \textbf{100}&\textbf{100}&68.42   &100 &100& 63.64   &   \textbf{3.41}&4.09&13.83  &3.41& 4.25& 14.00    \\

                           & GraphGAN   & \textbf{100}&\textbf{100}&57.89   &100& 96.57& 50.00 &    \textbf{3.94}&4.31&14.26  &4.00& 4.61& 14.50   \\

                           & LPA        & \textbf{100}&\textbf{100}&57.89   &100&100 &55.78    &    \textbf{3.75}&3.90&13.21          &3.82  &3.98&13.42      \\
                           & Louvain    & \textbf{100}&\textbf{100}&63.16   &100&98.54&58.86   &      \textbf{4.21}&\textbf{4.21}&12.53  &4.02&4.32  &13.86      \\ \cline{2-14}
                           & Average    & \textbf{100}&\textbf{100}&63.92   &100&99.02&59.29   &     \textbf{100}&\textbf{100}&63.92   &\textbf{100}&99.02&59.29   \\ \hline \hline
\end{tabular}
}
\end{table*}

\subsection{Limited Knowledge}

\label{sec:LD}
\begin{center}
\begin{figure}[!t]
\includegraphics[width=\linewidth]{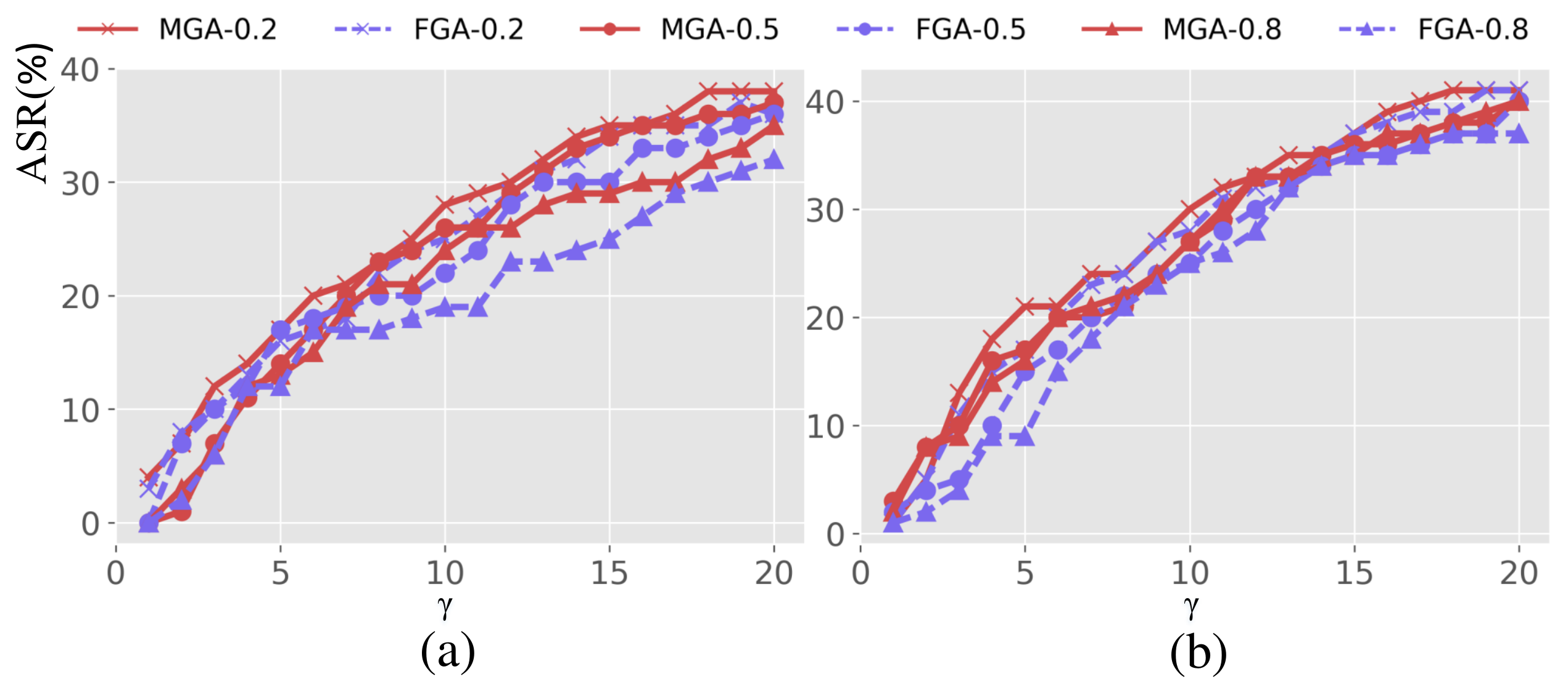}
\centering
\caption{The attack results of MGA and FGA when knowledge is limited. In each missing case, the probability of missing links is set to 0.2, 0.5, and 0.8. (a) Ensuring the 1-hop neighbors; (b) Randomly missing.}
\label{fig:Limite}
\end{figure}
\end{center}

In the previous experiments, we have assumed full knowledge of the input datasets, which may be unrealistic. In practice, only part information of the dataset is available. This raises the question of how MGA performs when knowledge is limited. In this experiment, we consider two cases of limited knowledge: the first is ensuring the 1-hop neighbors of the target node and the rest links are randomly missing, and the second is that all links are randomly. Due to it's easy to attack low-degree nodes, we randomly select 10\% nodes as the target nodes whose degrees are greater than the average value. Taking the Pol.Blogs dataset as an example, the effect of MGA is shown in Fig.~\ref{fig:Limite}.

From Fig.~\ref{fig:Limite}, we can see that MGA performs better than FGA when knowledge is limited, especially in the case of randomly missing. Besides, in the second case, the attack effect of both two methods is significantly better than that of the first case, suggesting that the target node is greatly affected by its neighbors. Moreover, we can find that as the missing probability increases, the attack effect of both methods decreases but is not significant. This may be because low-degree nodes are very vulnerable to attack, and only the target nodes with moderate degree values may have different results in different missing probability.

\section{Conclusion}
\label{sec:Conclusion}
In this paper, we propose a network adversarial attack algorithm MGA based on the momentum of gradient, which can fool network analysis methods and make the target nodes mis-analyzed, including node embedding, node classification and community detection. In this method, we first use the original network to train GCN model and extract the gradient of pairwise nodes based on the adversarial network. Then we calculate the momentum of gradient based on the gradients of previous iterations. By finding the the pair of nodes with maximum absolute momentum, we update the original network iteratively and generate the adversarial network. In the experiment we not only do uniform attack, hub-node attack and bridge-node attack on three network embedding methods, but also do community deception to to verify that MGA is not only useful for node classification attacks. Besides, We also consider the case of limited knowledge, and MGA is still successful.

\section*{Acknowledgment}
\label{sec:ackonw}
This research was supported by the National Natural Science Foundation of China under Grant No. 62072406, the Natural Science Foundation of Zhejiang Provincial under Grant No. LY19F020025, the Major Special Funding for "Science and Technology Innovation 2025" in Ningbo under Grant No. 2018B10063, the Special Scientific Research Fund of Basic Public Welfare Profession of Zhejiang Province No. LGF20F020016.

\ifCLASSOPTIONcaptionsoff
  \newpage
\fi




\bibliographystyle{IEEEtran}
\bibliography{refnew}
%




%


\vfill




\end{document}